\documentclass[aps,prb,preprint,groupedaddress,showpacs]{revtex4}
\usepackage[dvips]{graphicx}

\begin{document}

\title{Void-induced cross slip of screw dislocations in fcc copper}
\author{Takahiro Hatano}
\affiliation{Earthquake Research Institute, University of Tokyo, Tokyo 113-0032, Japan}
\author{Tetsuya Kaneko}
\author{Yousuke Abe}
\author{Hideki Matsui}
\affiliation{Institute for Materials Research, Tohoku University, Sendai 980-8577, Japan}

\date{\today}

\begin{abstract}
Pinning interaction between a screw dislocation and a void in fcc copper is investigated 
by means of molecular dynamics simulation.
A screw dislocation bows out to undergo depinning on the original glide plane at low temperatures, 
where the behavior of the depinning stress is consistent with that obtained by a continuum model.
If the temperature is higher than $300$ K, the motion of a screw dislocation is no longer restricted 
to a single glide plane due to cross slip on the void surface.
Several depinning mechanisms that involve multiple glide planes are found.
In particular, a depinning mechanism that produces an intrinsic prismatic loop is found.
We show that these complex depinning mechanisms significantly increase the depinning stress.
\end{abstract}

\maketitle

\section{introduction}
Plasticity of crystalline materials is mainly dominated by mobility of dislocations, 
which can move at much lower stress levels than theoretical strength of a perfect crystal.\cite{friedel}
Dislocations interact with other lattice defects such as voids or precipitates.
Such dislocation-obstacle interactions limit mobility of dislocations to result in hardening, 
and hence they are one of major concerns in materials science.
Until very recently, they were investigated exclusively by continuum models 
\cite{foreman,bacon,scattergood},
which involve a single glide plane and neglect atomistic discreteness.
Such continuum models are now developed to three-dimensional multi-dislocations model.\cite{zbib}
Although these attempts are promising, they must be suitably supplemented with insights regarding 
atomistic-scale phenomena, because pinning processes essentially involve core structures of dislocations.
Recent extensive {\it in-situ} experiments \cite{sarnek,matsukawa} have found that 
elementary processes of dislocation-obstacle interactions are more complicated than considered before, 
mainly due to their atomistic nature.

As an approach that is complemental to experiments, molecular dynamics (MD) simulations 
also begin to reveal remarkable dynamics of dislocation pinning.
For example, an edge dislocation absorbs vacancies when they interact with vacancy clusters.
\cite{osetsky1,wirth1}
Namely, edge dislocations can sweep volume defects.
This may be an elementary process of dislocation channeling, 
because the region in which a dislocation once glides contains less defects 
and subsequent dislocations can pass the region more easily.
Another example is the finding of a new bypass mechanism, which is quite different from the Orowan mechanism.
During the bypass mechanism, an edge dislocation that is pinned by an impenetrable precipitate 
bows out to form a screw dipole.
This screw dipole can cross slip, which results in a complex bypass process 
in which double cross slip plays an important role.\cite{hatano1}

The above two examples are good illustrations of usefulness of MD simulation in the field of dislocation physics.
While these studies involve edge dislocations, simulations on screw dislocations are also intriguing 
because they generally bear more complex nature than edge dislocations; a good example is cross slip.
Some promising attempts are already made\cite{rodney,osetsky2,osetsky3,nogaret,marian}, 
such as the interaction with a Frank loop or a precipitate.
Along the line of these studies, pinning dynamics of a screw dislocation in fcc copper is investigated in this paper.
We focus interaction between a screw dislocation and a void, 
because voids are ubiquitous defects in irradiated metals \cite{jumel}
and play an essential role in irradiation hardening.

We report several new pinning dynamics, in which a screw dislocation undergo double cross slip on the void surface 
so that the motion of a pinned dislocation is no longer restricted on a single glide plane.
Cross slip remains partial because of asymmetry in energetics of constriction.\cite{rasmussen}
This partial cross slip enriches the nature of pinning and generally inhibits depinning of dislocations, 
which results in the increases of the depinning stress.
It is also found that cross slip on the void surface is a thermally activated process 
because no such behavior is observed at $0$ K and $150$ K.
Comparison with a pinning dynamics of an edge dislocation \cite{hatano2} will be briefly discussed.

\section{a computational model}
Let us describe our MD model. We consider fcc copper utilizing a many-body interatomic potential 
of Finnis and Sinclair \cite{finnis} with the parameters being optimized for copper.\cite{ackland}
The lattice constant is $3.615$ \AA.
We adopt this model amongst the others because of its simplicity and computational efficiency.
The only drawback is underestimate of stacking fault energy, which is $36$ mJ/m$^2$, 
while it is estimated as $45-78$ mJ/m$^2$ in experiments.
This leads to somewhat wide stacking fault ribbon in simulation 
(approximately $6b$, where $b$ denotes the Burgers vector length), 
which does not seriously influence our results.\cite{note}

\begin{figure}
\includegraphics[scale=0.5]{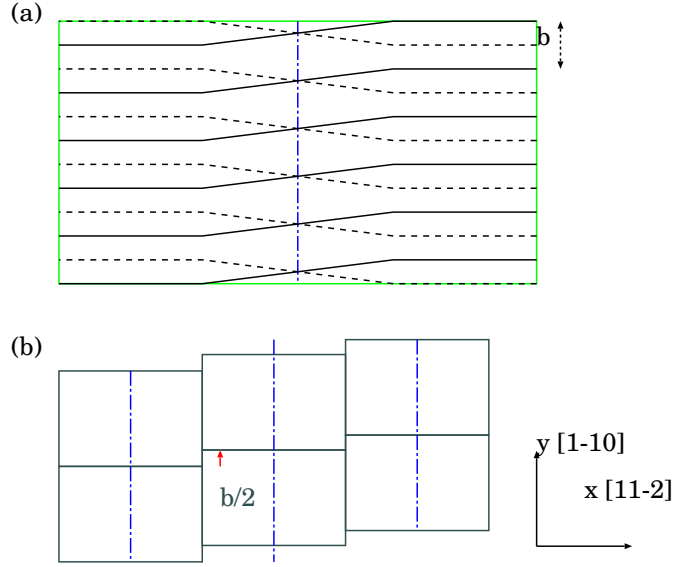}
\caption{\label{schematic}
(a) Schematic of the initial atomistic configuration for a screw dislocation.
The solid lines represent rows of atoms that belong to the region $z > 0$, 
while the dashed lines represent those in $z<0$.
The vertical dot-dashed line in the middle denotes a screw dislocation.
(b) Schematic of a boundary condition with respect to the $x$ direction.
Adjacent cells with respect to the $x$ direction are displaced by $b/2$ along the $y$ direction.}
\end{figure}

\begin{figure}
\includegraphics[scale=0.65]{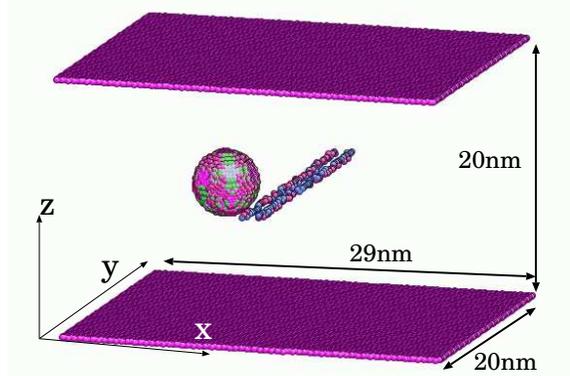}
\caption{\label{initial} (color online)
An initial configuration of the computational system where the void radius is $2$ nm here.
Initial distance between the void surface and the dislocation is $3$ nm.
Atoms are colored according to the number of the nearest neighbors \cite{li}: 
blue 13, red  11, green 10, and pink 9.
Atoms that have $12$ nearest neighbors, which constitute a perfect crystal, are omitted 
in order to visualize defects.}
\end{figure}

The dimensions of the model system are $29\times20\times20 {\rm nm}^3$.
This system consists of approximately $9.2\times 10^5$ atoms.
The $x$, $y$, and $z$ axes are taken along the $[11\bar{2}]$, $[1\bar{1}0]$, and $[111]$ directions, 
respectively.
In order to introduce a screw dislocation whose Burgers vector is parallel to the $y$ axis, 
we first prepare a perfect crystal and then displace atoms along the $y$ axis 
according to the following displacement fields.
\begin{equation}
\label{preparescrew}
u_y({\bf r})=\left\{
\begin{array}{@{\,}ll}
0, \ \ (x <0) \\
{\rm sign}(z) x/20, \ \ (0\le x \le 10b)\\
{\rm sign}(z) b/2, \ \ (x > 10b)
\end{array}
\right.
\end{equation}
where $b$ denotes the length of the Burgers vector. See Fig. \ref{schematic} for the schematic.
The displacement field $u_y$ yields a perfect screw dislocation whose strain field is symmetric 
with respect to $z=0$. \cite{note1}
Note that we cannot adopt a periodic boundary condition with respect to the $x$ direction, 
because the relative displacement of two adjacent cells is $b/2$.
Thus a modified periodic boundary condition is adopted, under which the adjacent cells are displaced by $b/2$ 
along the $y$ direction. See Fig. \ref{schematic}(b) for schematic.
As to the $y$ direction, an ordinary periodic boundary condition is employed 
so that we consider a screw dislocation of infinite length.
Note that the surfaces exist only in the $\pm z$ directions.
A void is introduced by removing atoms that belong to a spherical region whose radius is $r$.
Initial distance between the void surface and the dislocation is $3$ nm.
See Fig. \ref{initial} for the initial configuration explained above.

The velocities of copper atoms are given randomly according to the Maxwell-Boltzmann distribution.
We prepare systems at several temperatures: $10$, $150$, $300$, and $450$ K, respectively.
Then the system is relaxed without shear so that a screw dislocation introduced by Eq. (\ref{preparescrew}) 
dissociates into two Shockley partials separated by a stacking fault ribbon on $(111)$.
The distance between two partials is approximately $6b$.
Then the surfaces move antiparallel at a constant speed so that the system is subjected to plain shear.
We set the strain rate $\dot{\epsilon}=7 \times 10^6 \rm{s}^{-1}$.
Important parameters that we will vary in this study are the temperature and the void radius.
The shear stress is measured on the upper surface: the total force acting on the surface atoms divided by the area.

\section{behaviors of the depininng stress}
\begin{figure}
\includegraphics[scale=0.35]{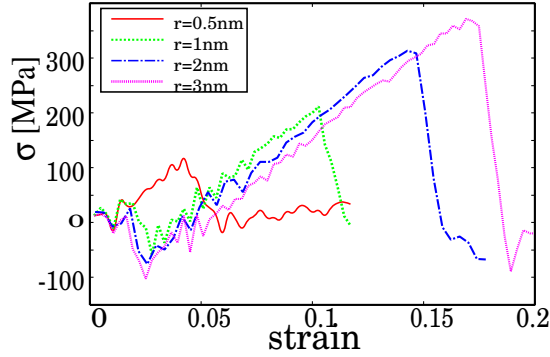}
\caption{(color online) The stress-strain relations in pinning processes for several voids at $10$ K.
The red solid line, the green dashed line, the blue dot-dashed line, and the pink dotted line 
represent those for $r=0.5$, $1$, $2$, and $3$ nm, respectively.}
\label{stressstrain}
\end{figure}

Stress-strain relations for pinning-depinning processes with voids of different radii 
are shown in Fig. \ref{stressstrain}.
Negative shear stress at the early stage is attributed to attractive interaction between a void and a dislocation.
Note that the stress-strain curve has a single peak, which means that two partial dislocations 
are depinned simultaneously.
This makes an apparent contrast to the case of edge dislocations \cite{hatano2}, 
where the stress-strain curve has two peaks, each of which corresponds to 
depinning of the leading and the trailing partials, respectively.
This is because the width of two partials of an edge dislocation (approximately $16b$) 
is much larger than that of a screw dislocation, $6b$.

\begin{figure}
\includegraphics[scale=0.27]{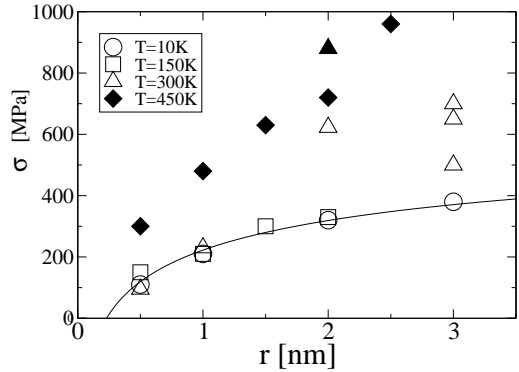}
\caption{\label{r-crss}
Void radius (\AA) dependence of the depinning stress $\sigma$ (MPa).
Different symbols denote the depinning stresses at different temperatures: 
The circles, the squares, the triangles, and the diamonds represent those 
at $10$, $150$, $300$, and $450$ K, respectively.
The solid line is Eq. (\ref{logarithmiclaw}) with $B=0.45$ nm.
At temperatures higher than $300$ K, the depinning stress does not obey Eq. (\ref{logarithmiclaw}) 
because the underlying mechanism changes.
The blank triangles at $r\ge 2$ nm  correspond to a depinning mechanism that involves double cross slip, 
while the solid symbols corresponds to a mechanism that yields an intrinsic prismatic loop 
(see section IV).}
\end{figure}

Then we discuss a quantitative aspect of the depinning stress.
At lower temperatures ($10$ and $150$ K), void size dependence of the depinning stress 
is described by a logarithmic law, which was originally discovered through a continuum model 
calculation.\cite{bacon}
\begin{equation}
\label{logarithmiclaw}
\sigma_{yz}=\frac{Gb}{2\pi L(1-\nu)}\ln\frac{1}{\left(0.5r^{-1}+L^{-1}\right)B}, 
\end{equation}
where $G$, $L$, and $\nu$ are the shear modulus, spacing between voids, and the Poisson ratio, respectively.
Note that $B$ is a parameter regarding the core-cutoff, which is on the same order of the Burgers vector length.
Although dissociation is not taken into account in this model, 
its quantitative validity is satisfactory as shown in Fig. \ref{r-crss}.
This is because the two partials are depinned simultaneously, as shown in FIG. \ref{stressstrain}.
In contrast, the depinning stress of edge dislocations cannot be described by Eq. (\ref{logarithmiclaw}) 
\cite{hatano2}, because two partials of an edge dislocation are widely separated 
so that the continuum model in which dissociation is not taken into account \cite{bacon} does not apply.
We confirm the depinning stress obeys Eq. (\ref{logarithmiclaw}) up to $T\le150$ K.

Although the behavior of the depinning stress at lower temperatures is reasonable as is discussed above, 
it does not obey Eq. (\ref{logarithmiclaw}) at temperatures higher than $300$ K.
Depinning stress is always greater than the prediction of Eq. (\ref{logarithmiclaw}).
Moreover, depinning stress is not uniquely determined by the configuration of the system, 
such as the void radius or the systems size.
It is not because of thermal fluctuation in the shear stress, 
but the complex (three dimensional) depinning mechanism of dislocations.
This depinning mechanism and its atomistic details constitute the main point of this study, 
which are discussed in the next section.

\section{void-induced cross slip}
\subsection{constriction on the void surface: the concepts of screwlike and edgelike constriction}
\begin{figure}
\includegraphics[scale=0.55]{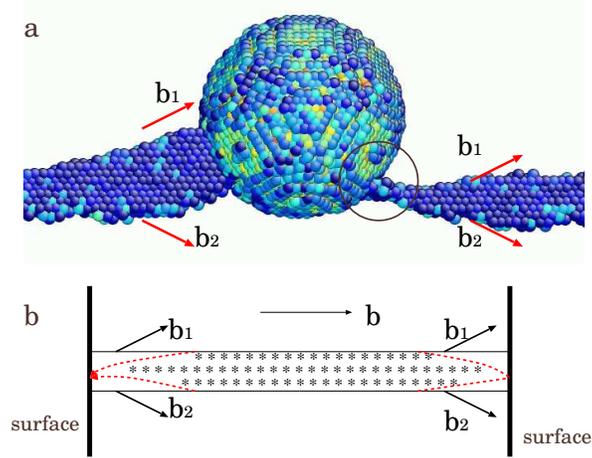}
\caption{\label{constriction}(color online)
(a) The $[111]$ projection of a void and a dislocation.
Atoms that constitute a perfect crystal are omitted, while those that constitute 
a dislocation and the stacking fault are visualized by the centrosymmetric parameter.
A dislocation undergoes constriction on the void surface, as is indicated by the circle.
Two vectors, $b_1$ and $b_2$, denote the Burgers vectors of two Shockley partials: 
$b_1=\frac{a}{6}[2\bar{1}\bar{1}]$ and $b_2=\frac{a}{6}[1\bar{2}1]$.
(b) A schematic figure of constriction on the surface.
The vector $b$ denotes the Burgers vector of a perfect screw dislocation, 
$\frac{a}{2}[1\bar{1}0]$.}
\end{figure}

Through careful observation of the atomistic configuration of the pinned dislocation, 
we find that cross slip occurs, which is preceded by constriction on the void surface.
The most important point is that constriction always occurs at one end of the dislocation, 
as is shown in Fig. \ref{constriction} a.
This asymmetry of constriction on the void surface is explained 
in terms of the energetics of Shockley partials.

As is shown in FIG. \ref{constriction} b, there are two different geometries 
with respect to the angles between their Burgers vectors and the normal vector of the surface.
These two cases have different energetics upon constriction.
On the left side of FIG. \ref{constriction} b, the angle between a partial dislocation and its Burgers vector 
decreases upon constriction so that the two partials are more like screw dislocations.
However, as is easily seen in FIG. \ref{constriction} b,  this tendency is quite opposite in the other configuration, 
where the angle between a partial dislocation and its Burgers vector increases upon constriction; 
i.e., the partials become edge-dislocation-like.
Following Rasmussen et al., we refer to constriction in the former configuration as screwlike constriction, 
while the latter as edgelike constriction.
Obviously, screwlike constriction is energetically more favorable than edgelike constriction.
Moreover, screwlike constriction has less energy than two parallel partials so that 
screwlike constriction is spontaneous on the free surface.
Indeed, Rasmussen et al. \cite{rasmussen} computed energy of these two geometries 
and found that the energy of screwlike constriction is $1.1$ eV lower than two parallel partials.
%Although thier calculation is based on an empirical interatomic pontetial, 
%more intuitive estimation can lead to the similar result as is explained below.
%First, we estimate the energy of the Shockley partials by decomposing them into screw part and edge part.

\subsection{constriction on the void surface leads to cross slip at high temperatures}
\begin{figure}
\includegraphics[scale=1.0]{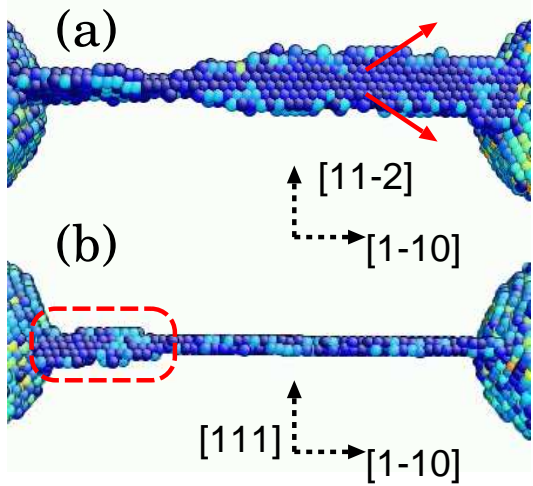}
\caption{\label{crossslip}(color online)
Cross slip occurs on the left side of a pinned dislocation via screwlike constriction.
On the right side, constriction is edgelike.
The visualization method is the same as that in FIG. \ref{constriction} a.
(a) The $[111]$ projection. The (red) thick arrows represents the Burgers vectors of partial dislocations.
(b) The $[11{\bar 2}]$ projection. Surrounded by the (red) dashed line is subjected to cross slip, 
which is on the $(11{\bar 1})$ plane.}
\end{figure}

As is explained above, constriction of the one end of a dislocation 
on the void surface (screwlike constriction) is spontaneous.
At $10$ and $150$ K, this spontaneous constriction does not lead to cross slip 
and the depinning process takes place exclusively on the $(111)$ plane.
However, at higher temperatures, constriction is followed by cross slip 
so that the partials move onto $(11{\bar 1})$ plane.
A typical atomistic configuration of cross slip on the void surface is shown in Fig. \ref{crossslip}.

Because constriction does not lead to cross slip at lower temperatures, 
it is speculated that cross slip requires some amount of activation energy 
even if dislocation is constricted on the surface.
This activation energy is explained in terms of the Burgers vectors of the Shockley partials.
If the constricted section of the dislocation cross slips, the Burgers vectors of the Shockley partials changes: 
$b_1$ changes from $a/6[2\bar{1}\bar{1}]$ to $a/6[2\bar{1}1]$ 
and $b_2$ changes from $a/6[1\bar{2}1]$ to $a/6[1\bar{2}\bar{1}]$.
(This situation is obvious when we recall the Thompson tetrahedron in which 
the right hand partial is Roman-Greek).
Due to this change, screwlike constriction on $(111)$ becomes edgelike 
if it cross slips onto $(11{\bar 1})$.
Eventually, energy difference between the two states is the activation energy of cross slip.

\section{three-dimensional depinning mechanisms}
In view of the concepts of the two types of constriction explained above, 
we discuss pinning dynamics of a screw dislocation in which cross slip plays an essential role.
Note that the discussions below involve higher temperatures at which 
constriction can lead to cross slip.
We find two different mechanisms, which are explained in the different subsections below.
The main difference between the two mechanisms is the occurrence of edgelike constriction.

\subsection{depinning mechanism A: double cross slip with edgelike constriction}
First we discuss the case that edgelike constriction occurs.
As is discussed in the previous section, the pinned dislocation undergoes cross slip 
via screwlike constriction, which is shown in FIG. \ref{crossslip}.
However, cross slip of the entire part of the dislocation is not possible without edgelike constriction, 
which costs a certain amount of energy 
($3.8$ eV according to Rasmussen et al.\cite{rasmussen}).
Therefore, until edgelike constriction occurs, cross slip remains partial 
and a dislocation is on the two glide planes simultaneously.
Due to the high activation energy, edgelike constriction takes place 
at relatively high shear stress, which is estimated as $150-200$ MPa in the present simulation.

If edgelike constriction is once realized at this stress level, 
it leads to cross slip of the entire part of a pinned dislocation 
onto the secondary glide plane, $(11{\bar 1})$.
Under the present configuration of simple shear, it is estimated that the shear stress 
on the secondary glide plane is $1/3$ that on the primary glide plane.
(Note that the factor $1/3$ is the Schmidt factor).
Thus depinning does not occur on the $(11{\bar 1})$ plane in the present simulation.
Instead, the dislocation double cross slips onto another $(111)$ plane and undergoes depinning there.
In this process, the depinning stress is somewhat (approximately $150$ \%) 
greater than Eq. (\ref{logarithmiclaw}), which is shown by blank triangles in FIG. \ref{r-crss}.

\subsection{depinning mechanism B: a prismatic loop formation}
\begin{figure}
\includegraphics[scale=0.75]{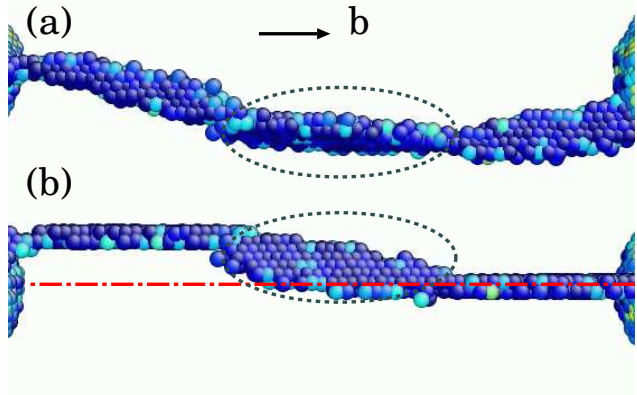}
\caption{\label{doublecrossslip}(color online)
A pinned dislocation undergoes double cross slip. Edgelike constriction does not take place.
(a) The $[111]$ projection of the atomistic configuration where partial double cross slip occurs 
on the left side via screwlike constriction.
The (red) dashed line indicates a region that undergoes double cross slip.
(b) The $[11{\bar 2}]$ projection of the same configuration as that in (a).
The dott-dashed line represents the original glide plane. The superjog is on the $(11{\bar 1})$ plane.
The visualization method is the same as those in the previous figures.}
\end{figure}

Then we discuss the case that edgelike constriction does not occur.
In this case, cross slip that is initiated by screwlike constriction does not propagate 
towards the other end of the dislocation.
Instead, cross-slipped section of the dislocation undergoes double cross slip 
onto another primary glide plane $(111)$, with the other end remaining on the original glide plane.
A typical atomistic configuration is shown in FIG. \ref{doublecrossslip}, 
where the dislocation is on two $(111)$ planes simultaneously.
Note that a superjog on the $(11{\bar 1})$ plane connects the two parts.
In this configuration, the dislocation cannot glide further and undergo complex depinning dynamics, 
which results in considerable increase of the depinning stress, as is explained in the next section.

\begin{figure}
\includegraphics[scale=0.75]{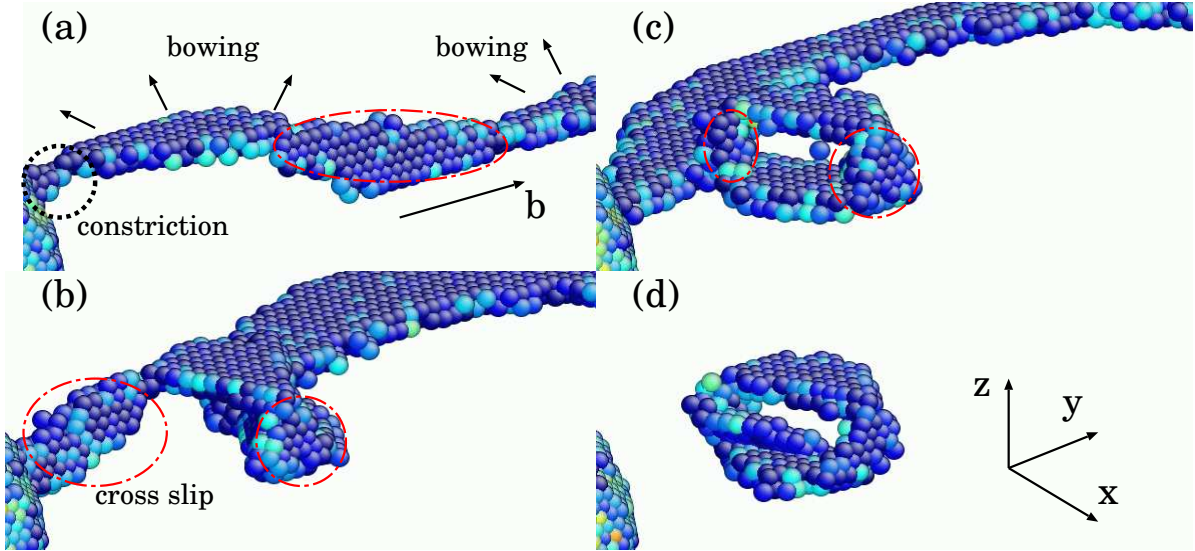}
\caption{\label{loopformation}(color online)
A depinning mechanism that yields an intrinsic prismatic loop.
The circles of red dot-dashed line indicates stacking fault ribbon on $(11\bar{1})$ planes.
(a) From the configuration shown in FIG. \ref{doublecrossslip}, screwlike constriction again occurs 
(indicated by the dotted circle).
(b) Screwlike constriction leads to cross slip.
The other segment is further bowing out, leaving stacking fault ribbon on $(11\bar{1})$, 
which is regarded as a superjog.
(c) The segment on the left side eventually double cross slips onto the original glide plane, $(111)$.
The twosegments recombine there to leave an intrinsic prismatic loop. 
(d) An intrinsic prismatic loop is left behind.}
\end{figure}

Note that the pinned dislocation is divided into two segments, each of which lies on 
the different $(111)$ planes.
While each segment bows out between voids and the superjog, 
the double-cross-slipped section eventually goes back onto the original glide plane 
via another double cross slip.
In this process, a dislocation loop is left behind upon depinning.
The atomistic configuration of this mechanism is shown in FIG. \ref{loopformation}.
By tracing the displacement of the atoms according to the Burgers vector, 
we find that this loop is an intrinsic prismatic loop.
As a result, the void is contracted.
In addition, it is obvious from FIG. \ref{loopformation} that the diameter of this prismatic loop 
is almost equivalent to the radius of the void. 

In this process, it is found that the depinning stress is approximately twice 
larger than Eq. (\ref{logarithmiclaw}). The reason is explained as follows.
Because the superjog can be regarded as a pinning point as well as the void, 
the spacing between pinning points $L$ is effectively half the original value 
when it is formed in the midsection of a pinned dislocation.
By replacing $L$ in Eq. (\ref{logarithmiclaw}) by $L/2$, we can expect the depinning stress is twice.
However, we remark that a superjog is not always formed in the midsection.
In such a case the depinning stress becomes much larger because the effective distance 
between the pinning points is less than $L/2$.
For example, in one case (the solid triangle shown in FIG. \ref{r-crss}), 
a superjog is formed in such a way that the distance between the void and the superjog is $L/3$.
As a result, the depinning stress is three times larger than the ordinary depinning stress 
described by Eq. (\ref{logarithmiclaw}).

\section{discussions and conclusions}
In this paper, we study pinning dynamics of a screw dislocation.
We find two novel depinning mechanisms, both of which involve cross slip on the void surface.
These pinning mechanisms are relevant to intermediate-high temperatures, 
because cross slip requires the activation energy.
The two mechanisms result in higher depinning stresses than previously predicted by a continuum model.
In the following, we remark three important points that are closely related to these mechanisms.

\subsection{Which mechanism to occur?}
Although we find two novel depinning mechanism, the following question still remains; 
which mechanism is dominant under a given condition?
From FIG. \ref{r-crss}, it is obvious that the prismatic loop formation process 
(i.e., mechanism B) is more likely at high temperature ($450$) K,  
whereas mechanism A is dominant at $300$ K.
Therefore, we can deduce that an additional activation process is essential to mechanism B.

In order to specify this process, we have to recall that mechanism B requires 
partial double cross slip before edgelike constriction occurs, 
the configuration of which is shown in FIG. \ref{doublecrossslip}.
Partial double cross slip is preceded by recombination of two partials on $(11\bar{1})$ 
and redissociation on $(111)$, which must be thermally activated.
Because the width of stacking fault ribbon on $(11\bar{1})$ is narrow due to the restricted 
geometry of the glide plane with respect to the void, this process is much easier than
edgelike constriction on$(111)$.

However, the precise energetics of this process is not straightforward.
The reason is that it occurs at the early stage of the pinning process, 
where the average shear stress is negative.
(See FIG. \ref{stressstrain} for negative shear stress at the early stage).
In such a situation spatiotemporal behavior of the shear stress near the void and the dislocation 
may be violent.

\subsection{Possibility of another but similar mechanism}
We remark another possible mechanism other than the two mechanism presented in section V.
Consider an intermediate stage of mechanism A, where the dislocation is on the $(11\bar{1})$ plane 
after edgelike constriction occurs.
Then bowing and depinning on the $(11{\bar 1})$ plane can occur under different shear loading.
For example, consider the case in which the maximum shear stress is applied to $\{100\}$ plane.
In this case, the shear stress is equally applied to four $\{111\}$ planes 
so that the depinning stresses with respect to these four glide planes are equivalent.
Thus the only barrier against depinning is edgelike constriction, 
which can be overcome by intermediate shear stress.
In such configurations, void can enhance cross slip and may play an important role 
in the nature of plasticity, particularly in work-hardening.
However, note that voids must be sufficiently large and temperatures must be high 
in order to realized cross slip via screwlike constriction.

\subsection{feasibility in experiments}
We remark that partial evidences for the existence of these phenomena are observed 
in experiments using transmission electron microscopy.\cite{nita}
In addition, because the shear rate is extremely high in MD simulations, 
an intrinsic prismatic loop that is produced by the present depinning mechanism 
may be somehow related to vacancy clusters that are frequently observed in high-speed deformation.
\cite{kiritani}
Further effort toward experimental validation of the depinning mechanisms proposed 
by the present MD simulation is in progress.

\acknowledgements
The authors gratefully acknowledges discussions with N. Nita on his experiments.


\begin{thebibliography}{99}
\bibitem{note}
As we will see in the later sections, constriction and cross slip are more difficult 
with wide stacking fault ribbon.
It implies that void-induced cross slip occurs more easily in experiments than in the present simulation.
\bibitem{note1}
This plane-symmetry of the initial strain field makes an apparent contrast to a screw dislocation
in the elastic body of infinite volume, whose strain field is cylindrically symmetric. But
this is not realizable in MD simulations because the computational cell is finite. Note that
this alternative approach is employed by several studies and seems to be reliable.\cite{rodney}
\end{thebibliography}
\end{document}